\documentstyle{article}
\begin{document}
\large
\begin{center}
{\LARGE A simple attacks strategy of BB84 protocol}\\[1.5cm]
{\large Guihua Zeng\footnote{Email: ghzeng@pub.xaonline.com} } \\[0.2cm]
{\small National Key Laboratory on ISDN of XiDian University, Xi'an, 710071, P.R.China\\[2.5cm]}
Abstract \\[0.2cm]
\end{center}

A simplified eavesdropping-strategy for BB84 protocol in quantum cryptography
is proposed. This scheme is based on the `indirect copying'. Under this scheme, eavesdropper can 
exactly obtain the exchanged information between the 
legitimate users without being detected.\\[0.2cm]
{\bf Key words}: Eavesdropping strategy, indirect copying, quantum cryptography, BB84 protocol, corresponding reference list.\\[1.0cm]

\begin{flushleft}
{\bf I.Introduction}
\end{flushleft}

Quantum cryptography, suggested originally by S.Wiesner [1] and then by C.H.Bennett 
and G.Brassard [2], employs
quantum phenomena such as the uncertainty principle and the quantum
corrections to protect distributions of cryptographic keys. Key distribution
is defined as procedure allowing two legitimate users of communication channel
to establish two exact copies, one copy for each user, of a random and secret
sequence of bits. In other words,
quantum cryptography is a technique that permits two
parties, who share no secret information initially, to communicate over an
open channel and to establish between themselves a shared secret sequence of
bits. Quantum cryptography is provably secure against eavesdropping attack,
in that, as a matter of fundamental principle, the secret data can not be
compromised unknowingly to the legitimate users of the channel.
BB84 protocol[3] is a key distribution protocol over an open channel by quantum 
phenomena, it relies on the uncertainty principle of quantum
mechanics to provide key security. The security guarantee is derived from
the fact that each bit of data is encoded at random on either one of a
conjugate pair of observables of quantum-mechanical object. Because such a
pair of observables is subjected to the Heisenberg uncertainty principle,
measuring one of the observables necessarily randomizes the other. 

Although quantum cryptography is provably security, with the quantum key distribution 
protocols presented, several attacks strategy have been generated, such as intercept/resend 
scheme [4], beamsplitting scheme [4], entanglement scheme [5-7] and quantum copying 
[9,10]. In the intercept/resend scheme, Eve intercepts selected light pulses and reads them 
in bases of her choosing. When this occurs, Eve fabricates and sends to Bob a pulse of 
the same polarization as she detected. However, due to uncertainty principle, at least 
25\% of the pulse Eve fabricates will yield the wrong result if later successfully measured 
by Bob. The other attack, beamsplitting, depends on the fact that transmitted light pulses 
are not pure single-photon states. In the entanglement scheme, the eavesdropper involves 
the carrier particle in an interaction with her own quantum system, referred to as probe, 
so that the particle and the probe are left in an entangled state, and a subsequent 
measurement of the probe yields information about the particle. Some investigators are 
now turning their attention to collective attacks and joint attacks. About these 
attacks description please see Ref.[8] and its references. Eve can also use the 
quantum copying to obtain the information between Alice and Bob. Two kind quantum copies are 
presented [10,11]. It is appropriate to emphasize the limitation of above attacks strategy.
All these mentioned attacks strategy are restricted by the uncertainty principle or the 
quantum corrections, so Eve can not break the quantum cryptography protocols.
The risk of eavesdropper is to disturb the information and finally to be detected by 
the legitimate users. This is the reason why quantum cryptography is declared to be 
provably security.

The Eve's aim is to obtain more information from the open channel set up by legitimates user, 
saying Alice and Bob, and induce more less disturbance on the transmitting quantum bits, 
so that she can not be detected by the legitimate users Alice and Bob. In usually, 
the uncertainty principle or the quantum corrections prevents Eve's attempt from 
eavesdropping the useful information without being detection. However if Eve can 
escape the restriction of the uncertainty principle, her attempt will be succeed. 

In this paper we  propose a novel attack strategy for quantum cryptographic protocols. 
Under this strategy, the security of BB84 quantum key distribution protocol will 
completely loss. The scheme works by follows procedure: Eve constructs a 
prescription function. This function must be an uniform function for every different
quantum state used by Alice and Bob. This mean that every function value corresponds to a
different quantum state. It consists of a reference list that all these corresponding relationship of function value to different quantum state. While Alice sends a random 
quantum sequence to Bob, Eve intercepts every state and calculates the corresponding 
value by the function, then gives up the intercepted
state. When this finishes, Eve resends a new quantum state to Bob according to the reference 
list in which every value corresponds a correct quantum state. By this method Eve can 
exactly obtain the information exchanged between Alice and Bob without being detected. 
We call this method as ``indirect copying". Of course,
It is different from the probabilistic cloning and the inaccurance quantum copying. 
Obviously, the ``indirect copying" is not a true copy of quantum bits.

\begin{flushleft}
{\bf II.BB84 quantum key distribution protocol}
\end{flushleft}

For describing our attacks strategy, we first review the BB84 protocol.
The BB84 protocol is the first key distribution protocol in quantum cryptography. 
Follows are the protocol in details [12].

\begin{enumerate}
\item Alice prepares a random  sequence of photons polarized and sends them to Bob
\item Bob measure his photon using a random sequence of bases
\item Results of Bob's measurements. Some photons are shown as not having been received owing to 
imperfect detector efficiency.
\item Bob tells Alice which basis he used for each photon he received.
\item Alice tells him which bases were correct.
\item Alice and Bob keep only the data from these correctly measured photons, discarding all the
rest.
\item This data is interpreted as a binary sequence according to the coding scheme:
\item Bob and Alice test their key by publicly choosing a random subset of bit positions and
verifying that this subset has the same parity in Bob's and Alice's versions of the key (here
parity is odd). If their keys had differed in one or more bit position, this test would have 
discovered that fact with probability 1/2.
\item Remaining secret key after Alice and Bob have discarded one bit from the chosen subset
in step 8, to compensate for the information leaked by revealing its parity. Step 9 and 10 are 
repeated k times with k independent random subsets, to certify with probability $1-2^{-k}$ that
Alice's and Bob's keys are the identical, at the cost of reducing the key length by k bits.
\item Distilling the security key by the privacy amplification [13].
The basic principle of
privacy amplification is as follows. Let Alice and Bob shared a random
variable $W$, such as a random $n$-bit string, while an eavesdropper Eve learns a
corrected random variable $V$, providing at most $t<n$ bits of information about
$W$, i.e., $H(W|V)\leq n-t$. Eve is allowed to specify an arbitrary distribution
$P_{VW}$ (unknown to Alice and Bob) subject to the only constraint that $R(W|
V=v)\leq n-t$ with high probability (over values $v$), where $R(W|V=v)$ denotes the
second-order conditional Renyi entropy of $W$, given $V=v$. For any $s<n-t$,
Alice and Bob can distill $r=n-t-s$ bits of the secret key $K=G(W)$ while keeping
Eve's information about $K$ exponentially small in $s$ , by publicly choosing the
compression function $G$ at random from a suitable class of maps into $\{0,1\}^
{n-t-s}$.
\end{enumerate}

\begin{flushleft}
{\bf III. Construction of Reference List}
\end{flushleft}

To perform privacy communication between legitimate users, known as Alice 
and Bob, a set of pre-defined nonorthogonal quantum states or noncommuting quantum states
often are used.  For briefly, We call this set of pre-defined nonorthogonal quantum state or
the noncommuting quantum states as basic quantum states (BQS) in the remainder paper. 
Because the BQS are publicly announced by Alice and Bob, Eve can easily get it.
In BB84 protocol, the BQS are the four noncommuting states $|0>, |\frac{\pi}{2}>, 
|\frac{\pi}{4}>, |\frac{3\pi}{4}>$. Of course the linearly polarized states $|0>, 
|\frac{\pi}{2}>$ and the circularly polarized states $|\frac{\pi}{4}>, |\frac{3\pi}{4}>$ are
orthogonal, respectively. 
In BB84 quantum key distribution protocol, the quantum states $|0>$ and $|\frac{\pi}{2}>$ are measured by the so called rectilinear measurement type. Representing this rectilinear 
measurement type as ${\cal L}$, we have
$${\cal L}|0>=\lambda_1 |0>, \eqno(1)$$
$${\cal L}|\frac{\pi}{2}>=\lambda_2 |\frac{\pi}{2}>, \eqno(2)$$
where $\lambda_i, i=1,2$ are eigenvalues. Because the states $|0>$ and $|\frac{\pi}{2}>$ constitute a base in Hilbert, an arbitrary quantum state can be expanded by this base, i.e.,
$$|\psi>=c_1|0>+c_2|\frac{\pi}{2}>. \eqno(3)$$
By Eq.(3), it is easy to obtain
$$|\frac{\pi}{4}>=\frac{\sqrt{2}}{2}|0>+\frac{\sqrt{2}}{2}|\frac{\pi}{2}>,\eqno(4)$$
$$|\frac{3\pi}{4}>=\frac{\sqrt{2}}{2}|0>-\frac{\sqrt{2}}{2}|\frac{\pi}{2}>,\eqno(5)$$
Consider a proper ancilla quantum state, for example,
$$|\alpha>=\frac{\sqrt{3}}{2}|0>+\frac{1}{2}|\frac{\pi}{2}>,\eqno(6)$$
making product between the ancilla quantum state $|\alpha>$ and the quantum of BQS gives
$$<\alpha|0>=\frac{\sqrt{3}}{2}\longrightarrow m_1=\frac{3}{4}=0.75, \eqno(7)$$
$$<\alpha|\frac{\pi}{2}>=\frac{1}{2}\longrightarrow m_2=\frac{1}{4}=0.25, \eqno(8)$$
$$<\alpha|\frac{\pi}{4}>=\frac{\sqrt{6}+\sqrt{2}}{4}\longrightarrow m_3=
\frac{(\sqrt{3}+1)^2}{8}\approx 0.933, \eqno(9)$$
$$<\alpha|\frac{3\pi}{4}>=\frac{\sqrt{6}-\sqrt{2}}{4}\longrightarrow m_4=\frac{(\sqrt{3}-1)^2}{8}\approx
0.067, \eqno(10)$$

Obviously, an observable value $m_j, j=1,2,3,4$  corresponds to only a basic quantum state
$|j_k>, k=1,2,3,4$.  All these 
corresponding relationship constructs a 
corresponding reference list. It is given by 
$$\begin{tabular}{|c|c|}
\hline
quantum state $|j_k>$& $m_k$\\
\hline
$|0>$ & 0.75\\
\hline
$|\frac{\pi}{2}>$ & 0.25\\
\hline
$|\frac{\pi}{4}>$ & 0.933\\
\hline
$|\frac{3\pi}{4}>$ & 0.067\\
\hline
\end{tabular}
\eqno(11)$$
List (11) constructs an uniform function between the sorting value and the BQS, i.e.,
$S_k=f(|j_k>)$, where $k=1,2,3,4, |j_k>$ represent the four basic quantum states. Obviously, 
$S_1\neq S_2\neq S_3\neq S_4$. When Alice connects Bob and exchanges 
information, Eve intercepts the sequences of 
the quantum bits. For each quantum bit in the sequence intercepted by Eve, she measures it
and obtains a corresponding sorting value. Comparing the sorting value to the reference 
list, Eve resends the corresponding quantum bit to Bob. For example, if the measurement 
value corresponds $m_2=1/4$, Eve resends the quantum state $|\frac{\pi}{2}>$ to Bob. Thus, Eve can exactly obtain 
the complete information exchanged between Alice and Bob, and escapes the detection of Alice 
and Bob. So under the presented attack strategy, the BB84 protocols is completely insecure.

\begin{flushleft}
{\bf IV. Attack scheme}
\end{flushleft}

First, Eve constructs a corresponding  reference list for every state of BQS.
For correctly determining the intercepted quantum states and resending the 
correct quantum bits to Bob, every basic quantum state $|j_k>$ must correspond to a different
reference value (marking the function value as $S_k, k=1,2,\cdots, m$). So Eve firstly need 
to construct an uniform function which is an one-to-one map of 
$|j_k>$ to the function value $m_k$. 

Second, Eve intercepts the random sequence of quantum states sent by Alice and calculates the
value for every intercepted state by measurement operation. For distributing the quantum key, Alice randomly choose the quantum state from the basic quantum 
state $|j_k>, k=1,2,\cdots, m$, and 
sends the randomly selected quantum bits sequence to Bob. The communication between Alice and 
Bob is in an open channel, which Eve can easily access. Eve intercepts the quantum bits 
sequences sent by Alice, and measures the observables $m_k, k=1,2,3,4$ for every
quantum bit. By the measurement values Eve calculates the corresponding sorting for every intercepted quantum state by her machine. 

Third, Eve gives up the intercepted quantum state. The Eve's operation  will limited by the uncertainty principle, her measurement disturbs the quantum state because she don't know  beforehand the every random quantum bit state. If Eve resends these intercepted states to Bob
like the Intercept/Resend attack strategy proposed in Ref.[6], she will reveal herself. To avoid 
these case, we let Eve give up all these intercepted states. 

Finally, Eve resends the corresponding quantum states. By the calculation sorting values 
obtained in step 2, Eve chooses a corresponding quantum bit state according to the reference 
list and resends it to Bob. The
resent quantum state is exactly same as that send by Alice, it seems
that Eve `copies' the Alice's quantum state. However, it is not a real copying, 
This `copying' is completely different from the probability and inaccurance copying. We call 
it as ``indirect copying".
By this method Eve can measure Alice's signal exactly, and resend an exact copy of it, thereby escaping detection. 

Our attack strategy makes the quantum cryptographic protocol at risk. Of course, our scheme can
not attack every protocol proposed previously. For example, we can not attack the Ekert  
protocol [14], because there is no information encoded there while the particle transits from the 
source to the legitimate users. In fact, our scheme is only valid for the protocol that quantum state is encoded in transit. Meanwhile, Eve must know the BQS. In addition, the interval time between two  adjacent quantum state of the resent quantum state should almost keep the same 
as that in Alice's random sequence of quantum bits so that Bob can not feel Eve.

\begin{flushleft}
{\bf V. Conclusion}
\end{flushleft}

In conclusion, we proposed an attack strategy  for BB84 key distribution protocol in the 
quantum cryptography, we called this strategy as `indirect copying attack'. Under this 
strategy, the BB84 quantum cryptographic protocols
is at risk, the eavesdropper can exactly obtain the information between the legitimate users 
without being detected. Of course, the presented strategy is only valid for the case that 
the quantum state is encoded in transit. 

\eject
\begin{enumerate}
\begin{center}
{\bf References}\\[0.1cm]
\end{center}
\item S. Wiesner, Conjugate coding, Sigact News, vol. 15, no. 1,78(1983); original manuscript written circa 1970. 
\item C.H.Bennett, G.Brassard, S.Breidbart and S.Wiesner, Quantum cryptography, or unforgeable subway tokens, Advances in cryptography: Proceedings of Crypto'82, August
1982, plenum, New York, pp. 267-275.
\item C.H.Bennett and G.Brassard. Quantum cryptography: Public key distribution and coin tossing, Proceedings of the IEEE International Conference on Computers, System, and Signal Processing, Bangalore, India (IEEE, New York. 1984), 
pp. 175-179.
\item C.H.Bennett, F.Bessett, G.Brassard, L.Salvail, and J.Smolin, Experimental quantum cryptography, J.Cryptology, 5, 
3(1992).
\item C.A.Fuchs, N.Gisin, R.B.Griffiths, C.S.Niu, and A.Peres, Optimal eavesdropping in quantum cryptography. I. Information bound and optimal strategy, Physical Review A 56, 1163 (1997).
\item R.B.Griffiths, and C.S.Niu, Optimal eavesdropping in quantum cryptography. II. A quantum circuit. Physical Review A, 56, 1173(1997).
\item C.H.Bennett, T.Mor, and J.A.Smolin, Parity bit in quantum cryptography, Physical Review A 54, 2675 (1996).
\item B.A.Slutsky, R.Rao, P.C.Sun, and Y.Fainman, Security of quantum cryptography against individual attacks, Physical Review A 57, 2383(1998).
\item M.Hilery, and V.Buzek, Quantum copying: Fundamental inequalities, Physical Review A 56, 1212(1997).
\item V.Buzek, and M.Hillery, Quantum copying: Beyond the no-cloning theorem, Physical Review A 54, 1844(1996).
\item L.M.Duan and G.C.Guo, Probabilistic cloning and identification of linearly independent
quantum states, Physical Review Letters 80, 4999(1998).
\item C.H.Bennett, Quantum cryptography using any two nonorthogonal states, Physical Review Letters, 68, 3121(1992).
\item C.H.Bennett, G.Brassard, C.Crepeau and U.M.Maurer, Generalized privacy amplification, IEEE Trans. Inform. Theory, 41, 1915(1995).
\item A.K.Ekert, Quantum cryptography based on Bell's theorem, Physical Review Letter, 67, 661(1991).
\end{enumerate}
\end{document}